\documentclass[twocolumn]{aastex62}

\usepackage{lineno}

\usepackage{newtxtext,newtxmath}
\usepackage[T1]{fontenc}
\usepackage{graphicx}	
\usepackage{amsmath}	

\usepackage{lineno}
\usepackage{xcolor}

\newcommand{\msun}{$M_\odot$}

\definecolor{darkgreen}{rgb}{0.1, 0.6, 0.1}

\begin{document}


\title[Quenched field dwarf galaxies]{The Environmental Quenching Mechanisms of Field Dwarf Galaxies}
\shorttitle{Origin of quenched field dwarf galaxies }
\shortauthors{J. A. Benavides et al.}

\author[0000-0003-1896-0424]{Jos\'e A. Benavides\href{https://orcid.org/0000-0003-1896-0424}{\includegraphics[scale=0.8]{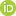}}}
\affiliation{Department of Physics and Astronomy, University of California, Riverside, CA, 92507, USA}

\author[0000-0003-3862-5076]{Julio F. Navarro\href{https://orcid.org/0000-0002-3790-720X}{\includegraphics[scale=0.8]{images/orcid.png}}}
\affiliation{Department of Physics and Astronomy, University of Victoria, Victoria, BC V8P 5C2, Canada}

\author[0000-0002-3790-720X]{Laura V. Sales\href{https://orcid.org/0000-0002-3790-720X}{\includegraphics[scale=0.8]{images/orcid.png}}}
\affiliation{Department of Physics and Astronomy, University of California, Riverside, CA, 92507, USA}

\author[0000-0003-1634-4628]{Isabel P\'erez\href{https://orcid.org/0000-0003-1634-4628}{\includegraphics[scale=0.8]{images/orcid.png}}}
\affiliation{Dpto. de F\'{\i}sica Te\'orica y del Cosmos, Facultad de Ciencias (Edificio Mecenas), University of Granada, E-18071, Granada, Spain}
\affiliation{Instituto Carlos I de F\'\i sica Te\'orica y Computacional, Universidad de Granada, E18071, Granada, Spain}

\author[0000-0002-2643-2472]{Bahar Bidaran\href{https://orcid.org/0000-0002-2643-2472}{\includegraphics[scale=0.8]{images/orcid.png}}}
\affiliation{Dpto. de F\'{\i}sica Te\'orica y del Cosmos, Facultad de Ciencias (Edificio Mecenas), University of Granada, E-18071, Granada, Spain}

\begin{abstract}
Field dwarf galaxies not actively forming stars are relatively rare in the local Universe, but are present in cosmological hydrodynamical simulations. We use the TNG50 simulation to investigate their origin and find that they all result from environmental effects that have removed or reduced their gas content.  Quenched field dwarfs consist of either backsplash objects ejected from a massive host or of systems that have lost their gas after crossing overdense regions such as filaments or sheets (``cosmic web stripping''). Quenched fractions  rise steeply with decreasing stellar mass, with quenched systems making up roughly $\sim 15\%$ of all field dwarfs (i.e., excluding satellites)  with stellar masses $10^{7}<M_{\star}/M_{\odot}<10^{9}$. This fraction drops to only $\sim1\%$ when a strict isolation criterion that requires no neighbours with $M_{\star}>10^9\, M_{\odot}$ within {$1.5$} Mpc is applied. Of these isolated dwarfs, $\sim 6\%$ are backsplash, while the other $\sim 94\%$ have been affected by the cosmic web. Backsplash systems are more deficient in dark matter, have retained less or no gas, and have stopped forming stars earlier than cosmic web-stripped systems. The discovery of deeply isolated dwarf galaxies which were quenched relatively recently would lend observational support to the prediction that the cosmic web is capable of inducing the cessation of star formation in dwarfs.
\end{abstract}


\section{Introduction}

Dwarf galaxies (defined here as systems with $M_{\star} < 10^9\, M_{\odot}$, or $M_r \gtrsim -17 $) are an important laboratory for studying the processes responsible for quenching star formation in galaxies. Unlike massive galaxies, where quenching is believed to be driven mainly by the feedback effect of central supermassive black holes, halting star formation in dwarf galaxies is thought to be due to feedback effects from their own stars, perhaps aided by environmental effects. At the very faint end (i.e., $M_{\star} < 10^5\, M_{\odot}$) cosmic reionization is also thought to play a substantial role \citep[see, e.g.,][and references therein]{Okamoto2009,Busha2010,Boylan-Kolchin2011,Benitez-Llambay2017,Benitez-Llambay2020,Pereira-Wilson2023}.\\

An important constraint on these ideas comes from the observation that essentially all field dwarfs (i.e., those that are not satellites of a massive host) with $10^7<M_{\star}/M_{\odot}<10^9$ are at present actively forming stars \citep{Geha2012}.  This has been taken to imply that quenching in dwarfs results from environmental effects able to remove fully or partially their gas content.\\

In this scenario, only dwarfs which have lost their gas through ram pressure \citep{GunnGott1972, Abadi1999} should be able to quench \citep[see e.g.,][]{Park2023}. Ram pressure is thus expected to affect only dwarf satellites; i.e., objects that are found today within the virial radius of a larger galaxy or galaxy system host. These hosts contain enough gas in their own halos to exert significant ram-pressure forces on the interstellar medium of the orbiting dwarf.\\

Ram-pressure stripping may also able to generate quenched galaxies in the field, through backsplash \citep{Balogh2000,Mamon2004}. \textit{Backplash galaxies} are, indeed, a population of objects found today in the field but which were in the past within the virial radius of a massive host, before being ejected to larger distances via gravitational interactions \citep[see e.g.,][]{Sales2007, Ludlow2009ApJ,Teyssier2012}.\\

Cosmological hydrodynamical simulations suggest a further quenching mechanism for field dwarfs, driven by interactions with the cosmic web, which may, under the right circumstances, strip much of the gas content of a dwarf. This mechanism, dubbed ``cosmic web stripping'' by \citet{Benitez-Llambay2013}, operates as  low-mass halos travel at high speed through the cosmic web and are affected by the ram pressure originating from the diffuse gas in the sheets and filaments of the web. This leads to the removal of low-density gas from a halo, reducing the gas reservoir available to sustain star formation. Note that this mechanism affects only the gas, and not the dark matter component of a system, and that it is most effective in low-mass halos, where gas is more weakly bound. Cosmic web stripping can operate even in low-density environments, far from the virial boundary of massive galaxies, and predicts the presence of a (small) number of truly isolated quenched galaxies. \\

Evidence of the impact of large-scale structures on the gas content of dwarf galaxies has also been explored in \citet{Herzog2023, Pasha2023, Samuel2023} using  cosmological simulations. These studies provide further evidence for a scenario where  the interaction of dwarf galaxies with their gaseous environments can lead to the suppression of star formation, and to the creation of a population of quenched dwarfs.\\

The environmental quenching mechanisms affecting field dwarfs (i.e., backsplash and cosmic web stripping) are a clear prediction of the hierarchical galaxy assembly process expected in the $\Lambda$CDM scenario and, therefere, observational confirmation is crucial. 
Tentative evidence in support of both quenching paths have been reported. For example, in the Local Group, the Cetus and Tucana dwarf spheroidals are candidate backsplash galaxies resulting from interactions with the MW \citep{Sales2007} or M31 \citep{Santos-Santos2023}. The dwarf irregular WLM has also been reported to show signs of ram-pressure stripping despite its relative isolation \citep{Yang2022}, suggesting cosmic web stripping as a possible culprit. Beyond the Local Group,  \citet{Stephenson2024} find evidence for environmental suppression of gas in low mass galaxies inhabiting filamentary structures.\\

However, the observational study of quenched dwarfs in the field is quite challenging, due to  their low luminosity and faint surface brightness. The pioneering study of \citet{Geha2012}, for example, is limited only to dwarfs down to masses exceeding $M_{\star} = 10^7\, M_{\odot}$. Extending such studies to fainter systems  is crucial to achieving a more complete understanding of dwarf quenching mechanisms in the Universe. Despite these difficulties, several such galaxies have been discovered serendipitously in the past decade \citep[e.g.,][]{Polzin2021,Prole2021,Sand2022,Casey2023, Carleton2024, Li2024}, and possibly even within cosmic voids \citep[e.g.,][]{roman2019,Bidaran2025}. However, a comprehensive large-scale study of field dwarfs and their star formation activity is still lacking. \\ 

While the focus of this work is on external mechanisms (primarily backsplash orbits and cosmic web stripping), internal processes such as supernova feedback, stellar winds, or even AGN activity can also significantly impact the star formation activity of low-mass galaxies. \citet{Chan2018}, for example, used zoom-in simulations to discuss the critical role of stellar feedback in regulating the star formation activity of isolated dwarf galaxies. This scenario, however, is not expected to lead to the full suppression of star formation in dwarfs, which is the main focus of our study.\\

In this paper, we use cosmological hydrodynamical simulations to investigate the mechanisms responsible for star formation quenching in field dwarfs. We focus on identifying the relative contributions of backsplash effects and cosmic web stripping, analyzing their respective signatures in the gas, stellar, and dark matter content of the dwarfs. By comparing these simulation predictions with observational constraints, we aim to provide a framework for differentiating between these quenching mechanisms in upcoming surveys, and to highlight the role of the cosmic web (in filaments or sheets) in shaping the evolutionary paths of star formation activity in dwarf galaxies. As we were finishing up this study, we became aware of a similar study by  \citet{Bhattacharyya2025}, which addresses similar topics but without our emphasis on identifying the specific quenching mechanisms affecting isolated field dwarfs . \\

The paper is organized as follows: In Sec.~\ref{SecSims} we briefly describe the simulation and some details of the sample of quenched field galaxies used in this work. In Sec.~\ref{SecResults} we present our analysis of the quenching process of field dwarfs. Finally, in Sec.~\ref{SecConc} we present and summarize our main results.\\

\begin{figure}
\centering
\includegraphics[width=\columnwidth]{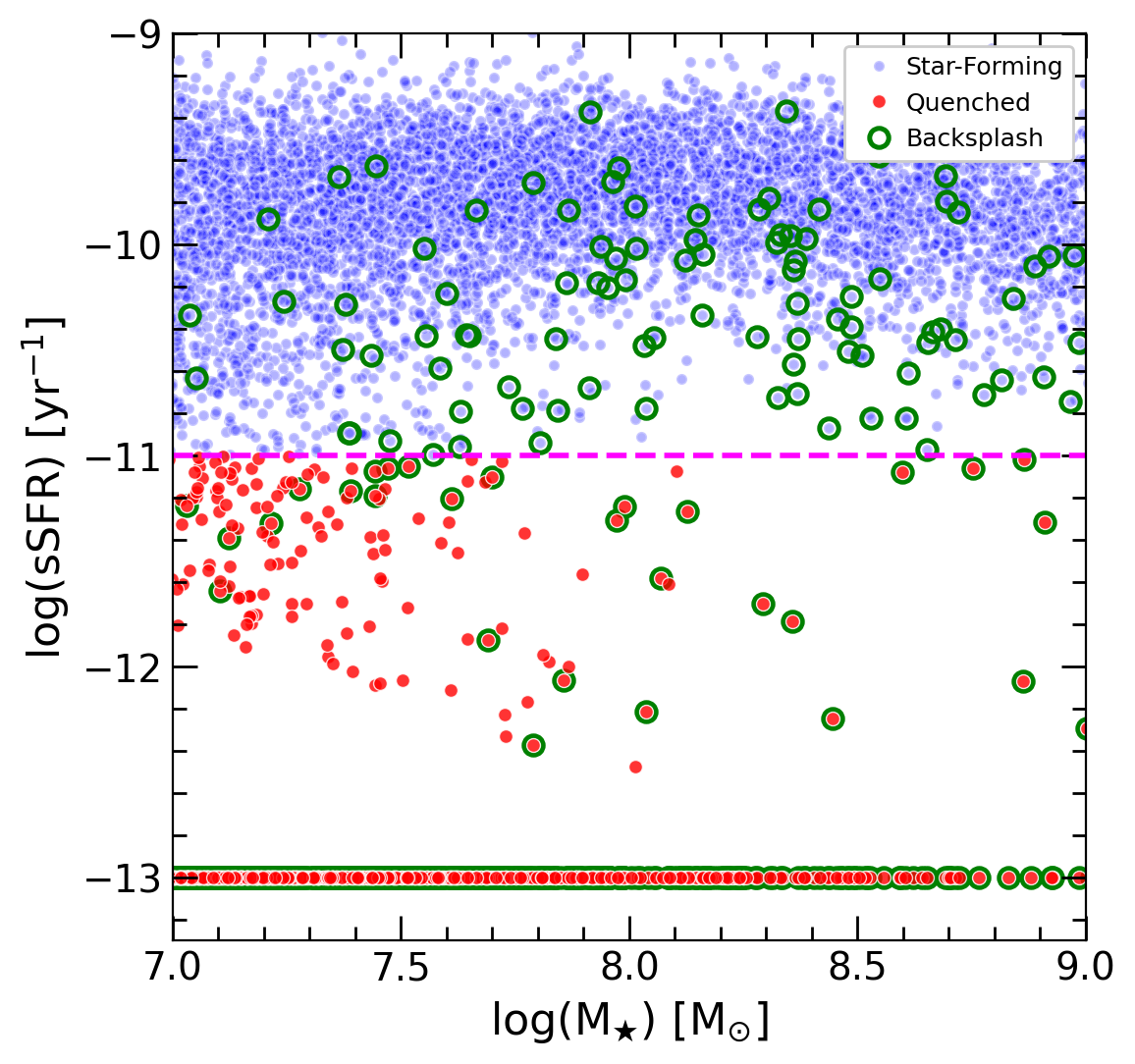}
\caption{Specific star formation rate (sSFR) as a function of stellar mass for all simulated central galaxies in the TNG50 simulation, split into star-forming (blue filled circles) and quenched (red filled circles) galaxies, following the criterion of \citet{Wetzel2012} at $\rm{sSFR = 10^{-11} ~ yr^{-1}}$ (dashed horizontal pink line). We also highlight the sample of backsplash galaxies (green open circles). Quenched galaxies with $\rm{SFR \sim 0}$ have been artificially placed at $\rm{sSFR = 10^{-13} ~ yr^{-1}}$.} 
\label{fig:SSFR}
\end{figure}

\section{Simulations and method}
\label{SecSims}

\subsection{The TNG50 simulation}
\label{ssec:simu}

In this study, we use the TNG50-1 (hereafter, TNG50 for short) cosmological hydrodynamical simulation \citep{Pillepich2018a, Pillepich2018b, Nelson2018, Pillepich2019, Nelson2019TNG}. TNG50 simulates the evolution of a cosmological volume approximately 50 Mpc on each side, using the {\sc arepo} code \citep{Springel2010}. The simulation adopts cosmological parameters  consistent with the measurements from \citet{PlankColaboration2016}\footnote{Cosmological constant $\Omega_{\Lambda} = 0.6911$, total matter content (dark matter + baryons) $\Omega_{\rm m} = \Omega_{\rm drk} + \Omega_{\rm b} = 0.3089 $, $\Omega_{\rm b} = 0.0486$, Hubble constant $H_0 = 100 \, h \, $km$ \, $s$^{-1} \, $Mpc$^{-1}$, $h = 0.6774$, $\sigma_8 = 0.8159$, and spectral index $n_s = 0.9667$}. The dark matter particle mass resolution is ${m_{\rm drk}} = 4.5 \times 10^5\, M_{\odot}$, while the typical baryonic element mass (gas cells or star particles) is $m_{\rm bar} \sim 8.5 \times 10^4\, M_{\odot}$. The gravitational softening for both dark matter and stars is set to $\epsilon = 0.29$ kpc at $z=0$, with the softening becoming significantly smaller for gas cells in high-density regions, potentially reaching $50\, h^{-1}$ pc.

\subsection{Simulated galaxies}
\label{SecSimGals}

We use the friends-of-friends \citep[FoF;][]{Davis1985} and {\sc subfind} \citep{Springel2001, Dolag2009} group finding algorithms to identify bound structures (halos and subhalos) in the simulation. Object evolution over time is tracked using the SubLink merger trees \citep{RodriguezGomez2015}. Virial quantities, such as mass, radius, and velocity ($\rm{M_{200}}$, $\rm{r_{200}}$, and $\rm{V_{200}}$ respectively) are determined at the radius where the average density equals $200$ times the critical density of the Universe ($\rho_{\rm crit} = 3 H^2 / 8 \pi G$).\\

In each FoF halo {\sc subfind}  identifies a ``central'' galaxy and ``satellite'' galaxies that are associated to it. The merger trees are able to track central galaxies which were, in the past, satellites of other massive hosts, but are at present outside the virial boundary of the host today. We shall hereafter refer to these galaxies as ``backsplash'' \citep[see e.g.,][]{Mamon2004,Muriel2014, Benavides2021}.\\

Galaxy properties such as stellar mass,  gas mass, and star-formation rate, are calculated inside a  ``galactic radius'' ($r_{\rm gal}$),  defined as twice the half-mass radius of the stars (${r_{h,\star}}$). Given the star particle mass of TNG50, we set a minimum stellar mass of $M_{\star}=10^{7}\, M_\odot$ in our analysis in order to resolve each galaxy with at least $\sim 120$ star particles. Our study thus shall consider mainly systems with stellar masses $10^{7}<M_{\star}/M_{\odot}<10^9$ at $z=0$. We shall only consider only central galaxies in this study and will exclude from the analysis any system labeled as a satellite by {\sc subfind}. We use the shorthand ``field dwarf'' to refer to any central galaxy with $10^{7}<M_{\star}/M_{\odot}<10^9$ at $z=0$. \\

In TNG50, the instantaneous star formation is also recorded for each simulated galaxy at all times. We define as ``quenched'' those galaxies whose specific star formation rate (sSFR $={\dot M_{\star}}/M_{\star}$) is below $< 10^{-11}\, $yr$^{-1}$, as in \citet{Wetzel2012}.\\

\section{Results}
\label{SecResults}

\subsection{Sample of quenched dwarf galaxies}
\label{ssec:sample_dwarfs}

Fig.~\ref{fig:SSFR} shows the specific star formation (sSFR) as a function of stellar mass for all field galaxies in TNG50 in the mass range of interest: $M_{\star}=[10^7$-$10^{9}] ~ M_{\odot}$. Objects with no star formation today have been all arbitrarily placed at $\rm sSFR = 10^{-13} \; \rm yr^{-1}$ for visualization purposes. This figure clearly highlights the presence of two subpopulation of field dwarfs: a ``star-forming" sample (blue) and a ``quenched" sample (red). We flag with green open circles all backsplash objects as identified by the merger trees.\\

\begin{figure}
	\centering
	\includegraphics[width=\columnwidth]{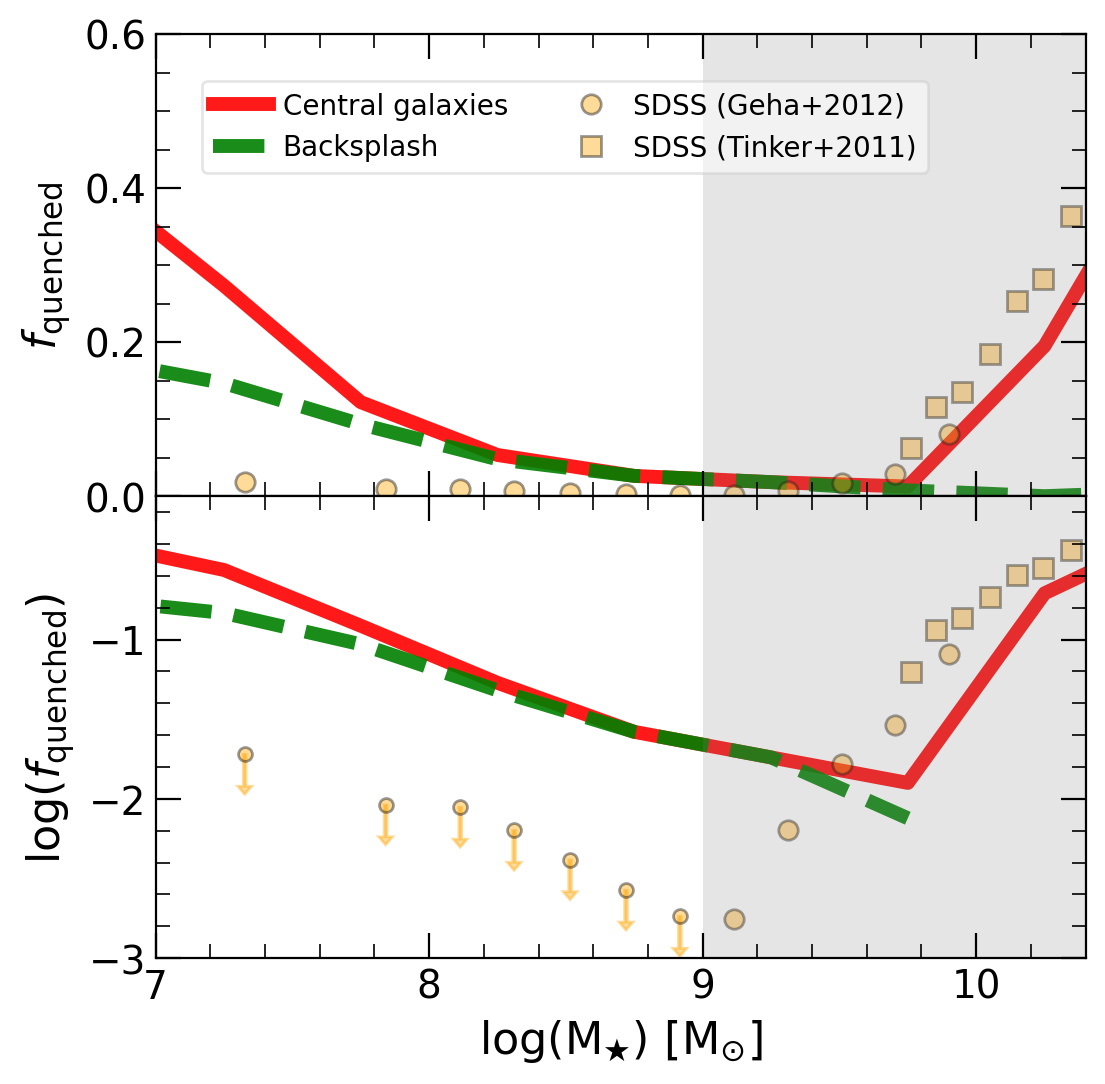}
	\caption{Fraction of quenched galaxies as a function of stellar mass. The solid red line denotes all simulated central galaxies in TNG50, while the dashed green line corresponds to backsplash galaxies. For comparison, observational data of quenched isolated galaxies from SDSS are shown using orange symbols \citep{Tinker2011, Geha2012}. The focus of this work is on low-mass galaxies with stellar masses in the range $M_{\star} = 10^7 - 10^9 ~ M_{\odot}$, below the shaded vertical region. The top panel presents the y-axis on a linear scale, while the bottom panel uses a log scale.}
	\label{fig:frac_quench}
\end{figure}

Backsplash systems make up a fair fraction of all quenched field dwarfs, with some interesting trends as a function of stellar mass. Essentially $100\%$ of massive quenched dwarfs are backsplash, indicating that encounters with more massive systems in the past are responsible for quenching dwarfs at the massive end. However, for less massive objects, many quenched systems are not backsplash, implying an alternative origin for their lack of star formation today.\\

\begin{figure*}
	\centering
	\includegraphics[width=1.0\textwidth]{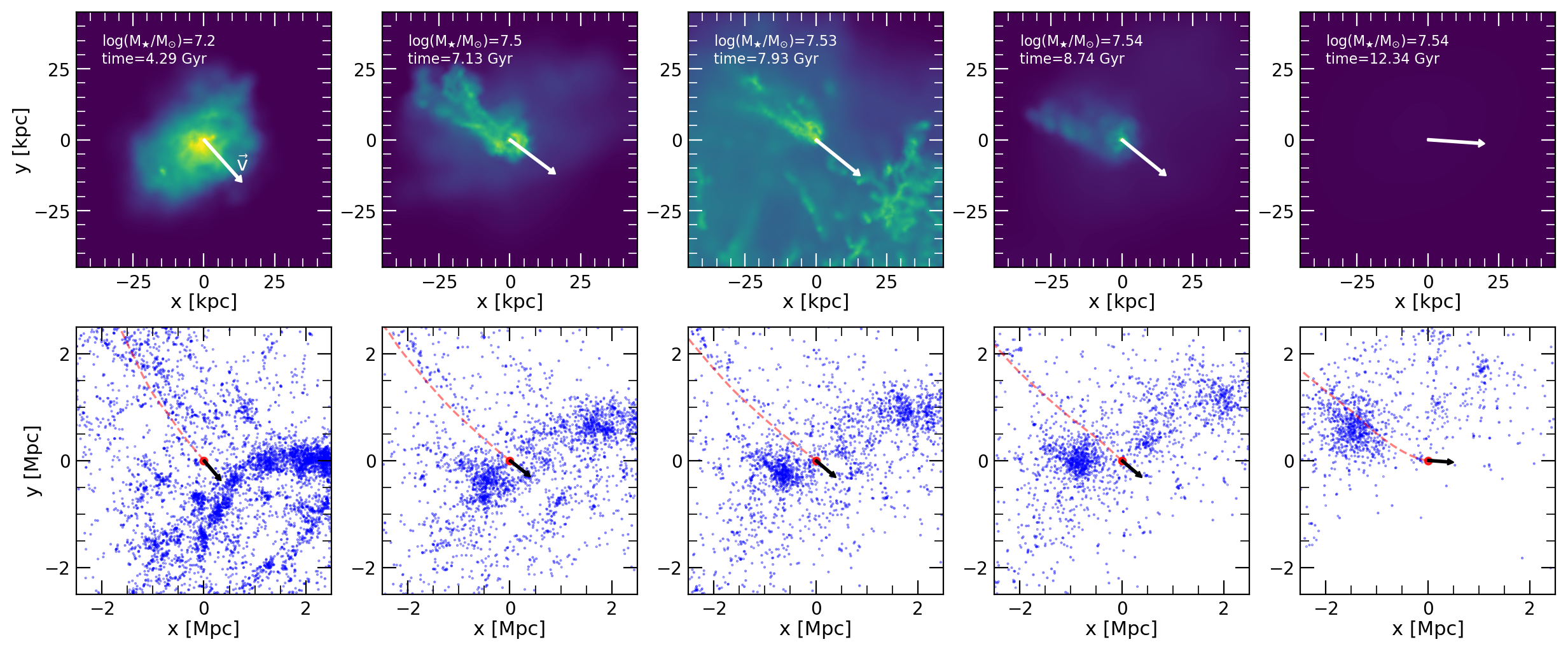}
	\includegraphics[width=1.0\textwidth]{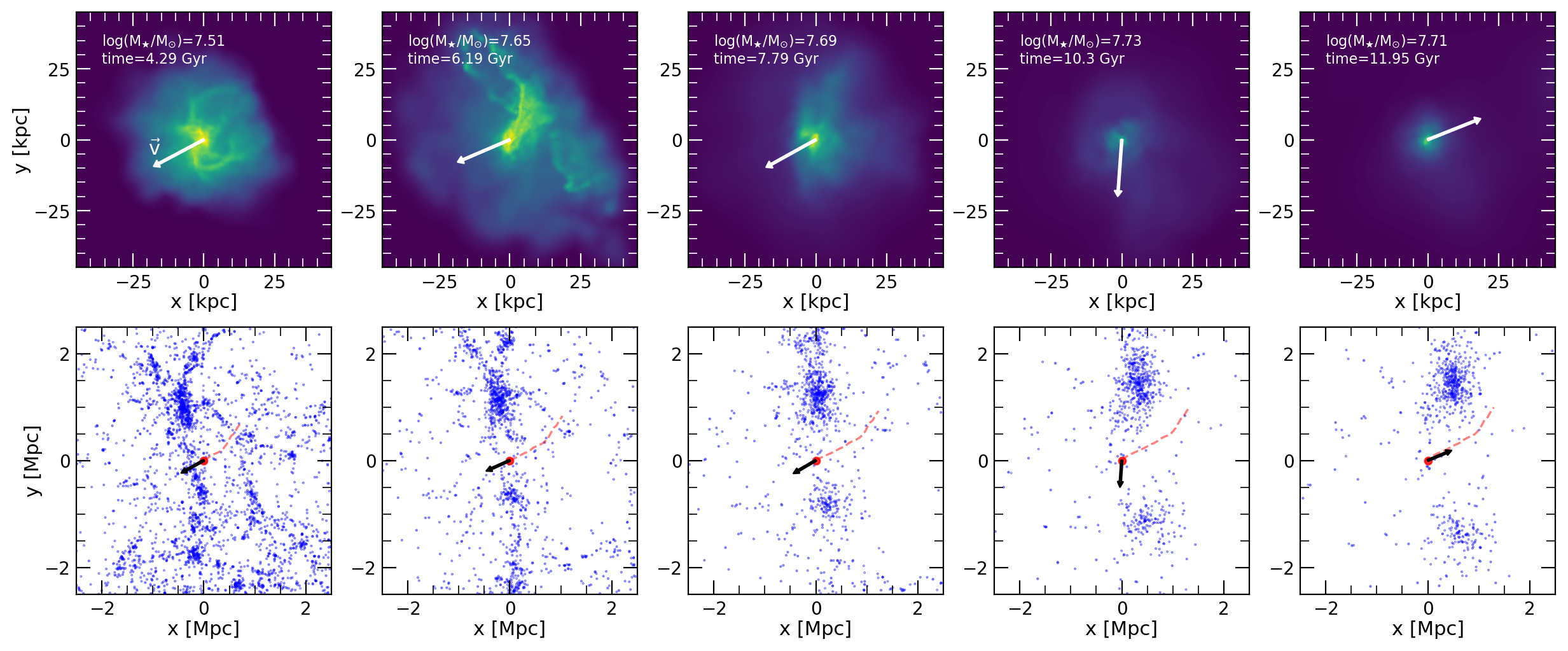}
	\caption{Spatial distribution of the gas content (first and third rows) in specific snapshots showing the evolution of two examples of isolated dwarf galaxies quenched through cosmic web stripping \citep{Benitez-Llambay2013}, and the large-scale environment (second and fourth rows) with the orbit followed by each galaxy over time, indicated by the thin red line. In all cases, the arrows represent the direction of the instantaneous velocity vector ($\bar{v}$) of the galaxies, relative to the simulated cosmological box coordinates. In both examples, the galaxies are quenched due to cosmic web stripping, which removes almost all of their gas content (completely in the first case a few Gyr after crossing the high-density gas region).}
	\label{fig:ex2_interaction}
\end{figure*}

\begin{figure*}
	\centering
	\includegraphics[width=\columnwidth]{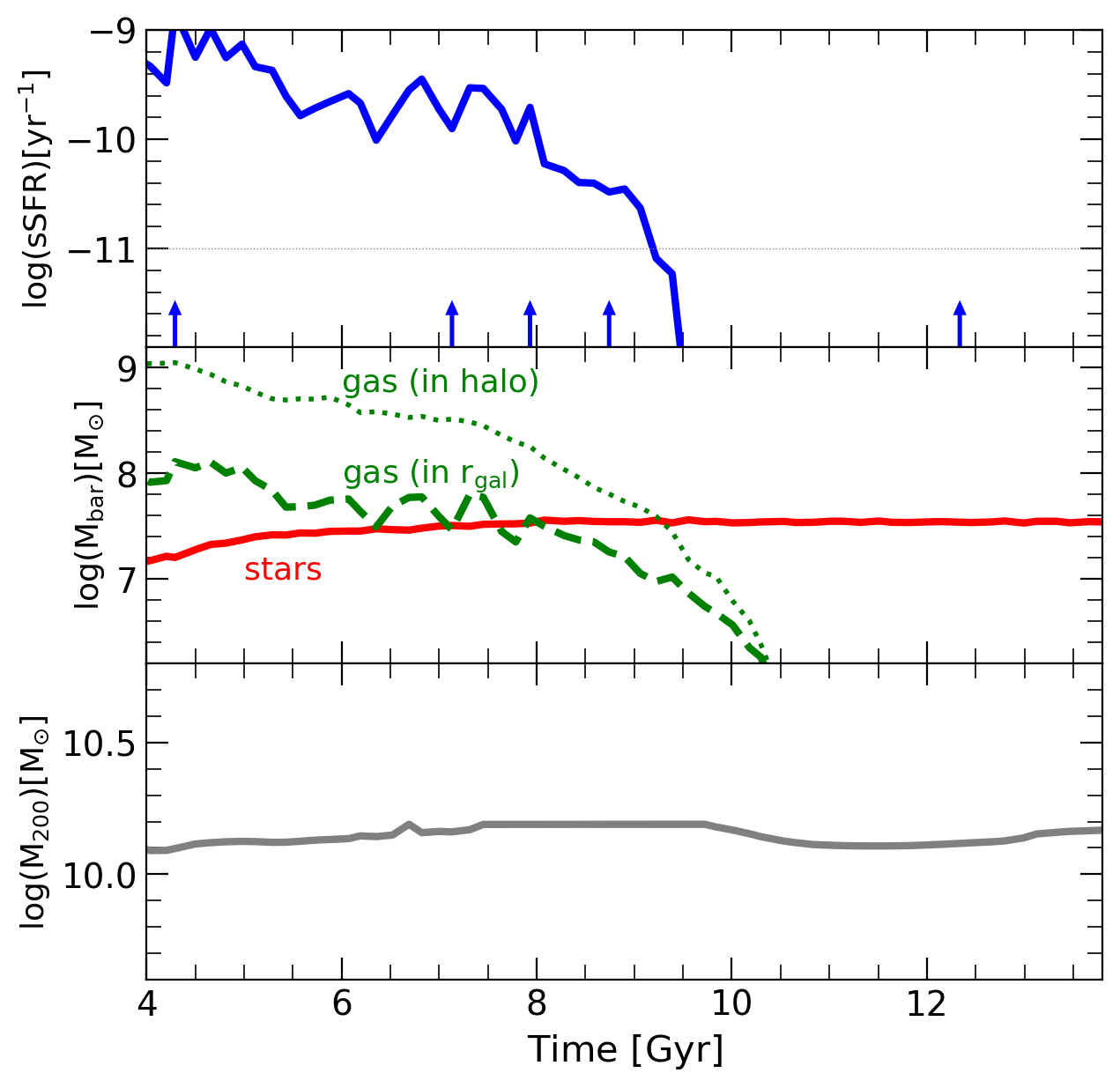}
	\includegraphics[width=\columnwidth]{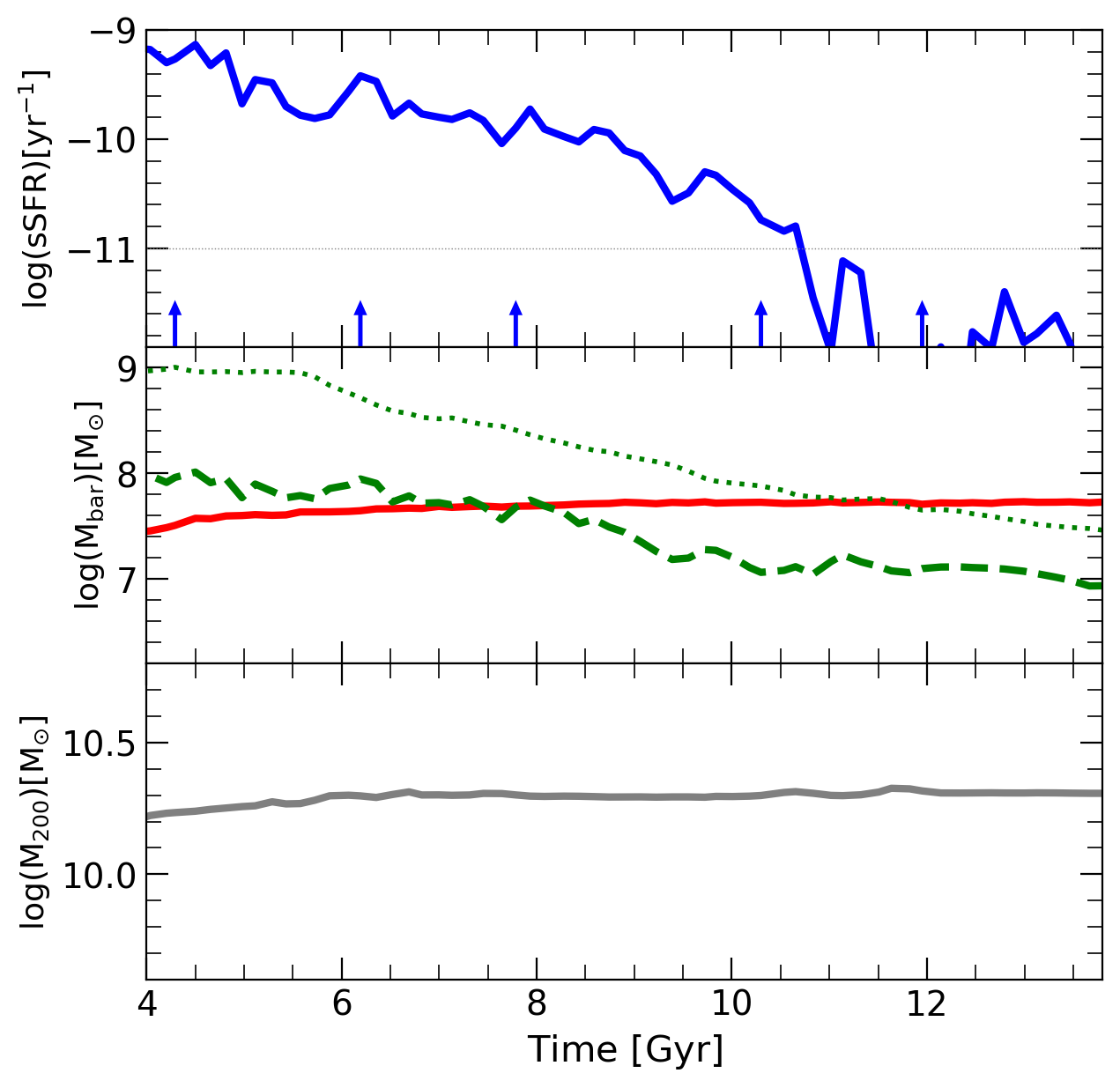}
	\caption{Examples of the evolution parameters for the same two isolated quenched galaxies showed in Fig.~\ref{fig:ex2_interaction}. The top panels show the specific star formation rate (sSFR) in blue, with the threshold for quenching criteria  ($\rm{sSFR/yr^{-1} = 10^{-11}}$) indicated by the grey horizontal line. The vertical blue arrows indicate the snapshots shown in Fig.~\ref{fig:ex2_interaction}. In the middle panels, the evolution of baryonic components, stars and gas within $\rm{r_{gal}}$ are represented by red solid line and green dashed line, respectively, while the total gas in the halo is shown by the green dotted line. The bottom panels show the evolution of the halo mass, represented by the grey line.}
	\label{fig:evol_ex1}
\end{figure*}

We explore this in more detail in Fig.~\ref{fig:frac_quench}, where we show the fraction of galaxies that are quenched as a function of stellar mass (solid red line). Note that the results shown include a range of masses wider than the dwarf regime studied here, mainly to put the population of low mass galaxies into context with other, more massive, counterparts. The fraction of quenched galaxies steadily decreases with stellar mass, from $\sim 30\%$ for centrals with $M_{\star} = 10^{10.5} ~ M_{\odot}$, to  only a few percent at the scale of $M_{\star} \sim 10^{9.7} ~ M_{\odot}$. Surprisingly, this decreasing trend with smaller $M_{\star}$ reverses for less massive dwarfs ($M_{\star} \lesssim 10^9 ~ M_{\odot}$), which show an increase in quenching fraction that starts at $M_{\star} < 10^8 ~ M_{\odot}$ and (becomes steeper) reaches $\sim 35\%$ at $M_{\star} \sim 10^7 ~ M_{\odot}$.\\

Observational studies of field quenched galaxies using the Sloan Digital Sky Survey (SDSS) are added in Fig.~\ref{fig:frac_quench} for comparison. In particular, we include results presented in \citet{Tinker2011} (orange squares) and in \citet{Geha2012} (orange circles). In observations too, the quenched fraction is seen to be relatively high at large galaxy masses, and to decline steeply with decreasing galaxy mass, a behaviour that has been interpreted as reflecting the increasing importance of AGN-feedback in regulating the star formation activity in more massive galaxies.\\ 

Encouragingly, the simulated results track the decreasing trend in quenched fraction when moving from $M_{\star} \sim 10^{10} ~ M_{\odot}$ to dwarfs with $M_{\star} \sim 10^{9} ~ M_{\odot}$. At smaller masses, however, the simulation shows an upturn in the quenched fraction so that the quenched fraction at $M_{\star} \sim 10^{7} ~ M_{\odot}$ matches that of $M_{\star} \sim 10^{10} ~ M_{\odot}$ galaxies. Interestingly, the upper bounds reported in \citet{Geha2012} do not rule out a slight increase in the quenched fractions after a minimum at $M_{\star} \sim 10^{9} ~ M_{\odot}$, but the incidence of quenching in the SDSS sample seems altogether smaller than in TNG50. At the same time, we note that the isolation criteria used in simulation and observations differ, so a precise match would be unexpected.\\

The main take-away point from Fig.~\ref{fig:frac_quench} is, instead, that the TNG50 simulation predicts robustly that the quenched fraction should increase towards lower stellar masses, after a minimum at $M_{\star} \sim 10^{9}\,$\msun, a prediction that should be testable observationally. Understanding the reason for this increase involves exploring the quenching mechanisms which affect these objects, an issue we turn to next.\\ 

As discussed in the Introduction, backsplash orbits are thought to be responsible for the existence of quenched galaxies in the field. We therefore split in Fig.~\ref{fig:frac_quench} the contribution of backsplash objects to the quenched galaxy fraction (green dashed line), finding that, while backsplash orbits account for all quenched dwarfs in the range $M_{\star}=10^8 - 10^9$\msun, lower mass dwarfs show the need for an alternative quenching mechanism that leads to the cessation of star formation in nearly half of quenched dwarfs at $M_{\star}=10^7\, M_{\odot}$\\

Careful examination of the non-backsplash cases reveals that cosmic web stripping \citep[e.g.,][]{Benitez-Llambay2013, Herzog2023} is the additional mechanism responsible for  quenching low-mass field dwarfs. Fig.~\ref{fig:ex2_interaction} shows the evolution of two systems chosen to illustrate this mechanism. In this figure, each row presents a set of snapshots, chosen to highlight the main event that led to the quenching of the system. The first row shows the gas distribution around one of the dwarfs; the second row shows, at the same times, its motion across the cosmological box (indicated by the dashed line). In the latter, the blue dots indicate the position of other galaxies in TNG50, in a cubic box centered on the dwarf.\\

The bottom two rows depict the same, but for a different system. In each of the panels, the arrows (white or black) indicate the direction of the galaxy peculiar velocity in the reference rest frame of the full cosmological box. The evolution of the star formation rate, baryonic mass, and halo mass of these two objects are shown in Fig.~\ref{fig:evol_ex1}.  \\

The examples shown in Fig.~\ref{fig:ex2_interaction} highlight how cosmic web stripping occurs in these systems, leading to their quenching. Two dwarfs of similar mass, $M_{\star} \sim 5 \times 10^7 \, M_{\odot}$, cross a filament-like region where the ambient gas ram-pressure strips much of the gas content of the dwarf \citep{Benitez-Llambay2013}. The tails of stripped gas extending in the direction opposite to the velocity of each galaxy (little arrows) is the smoking-gun signature of ram-pressure stripping taking place.\\

The interaction with the diffuse gas in the filament leads to the removal of a significant fraction of the gas from the dwarfs, a process that is more pronounced in the case of the dwarf shown in the top two rows of Fig.~\ref{fig:ex2_interaction}. Cosmic web stripping is responsible for removing much of the gas reservoir from the dwarf before fully quenching it by $t \sim 9.5$ Gyr. In the case of the second dwarf,  the removal of gas (and its effect on star formation) is more gradual; this system becomes fully quenched only by $t \sim 10.5$ Gyr (see top rows in Fig.~\ref{fig:evol_ex1}). The bottom panels in Fig.~\ref{fig:evol_ex1} show, as expected, that cosmic web stripping has little to no impact on the dark matter content of the dwarfs.\\

We emphasize that cosmic web stripping affects mainly systems at the low mass end; indeed, the large majority of non-backsplash quenched dwarfs have $M_{\star}<10^8 ~ M_\odot$. We expect this mechanism to become more prevalent in lower mass systems, where the lower potential depth would facilitate the removal of gas via these environmental mechanisms. The detection of a population of quenched, isolated dwarfs would provide strong support for the idea that the cosmic web is able to modulate the star formation of isolated dwarfs.\\

\subsection{Properties of quenched isolated dwarfs}
\label{ssec:quench_dwarfs}
As discussed earlier, the quenching of backsplash galaxies is relatively simple to understand, as most dwarfs  that venture into the virial boundary of a more massive system are likely to have their gas ram-pressure stripped by the gaseous atmosphere of the host. These systems are also likely to lose dark matter in the process, through tidal stripping. Backsplash systems are therefore expected to be outliers in the scaling relations between stellar mass, gas mass, and halo mass. In the case of cosmic-web stripping, tidal effects are relatively minor, and should not lead to significant removal of mass in the interacting halos. \\

\begin{figure}
	\centering
	\includegraphics[width=\columnwidth]{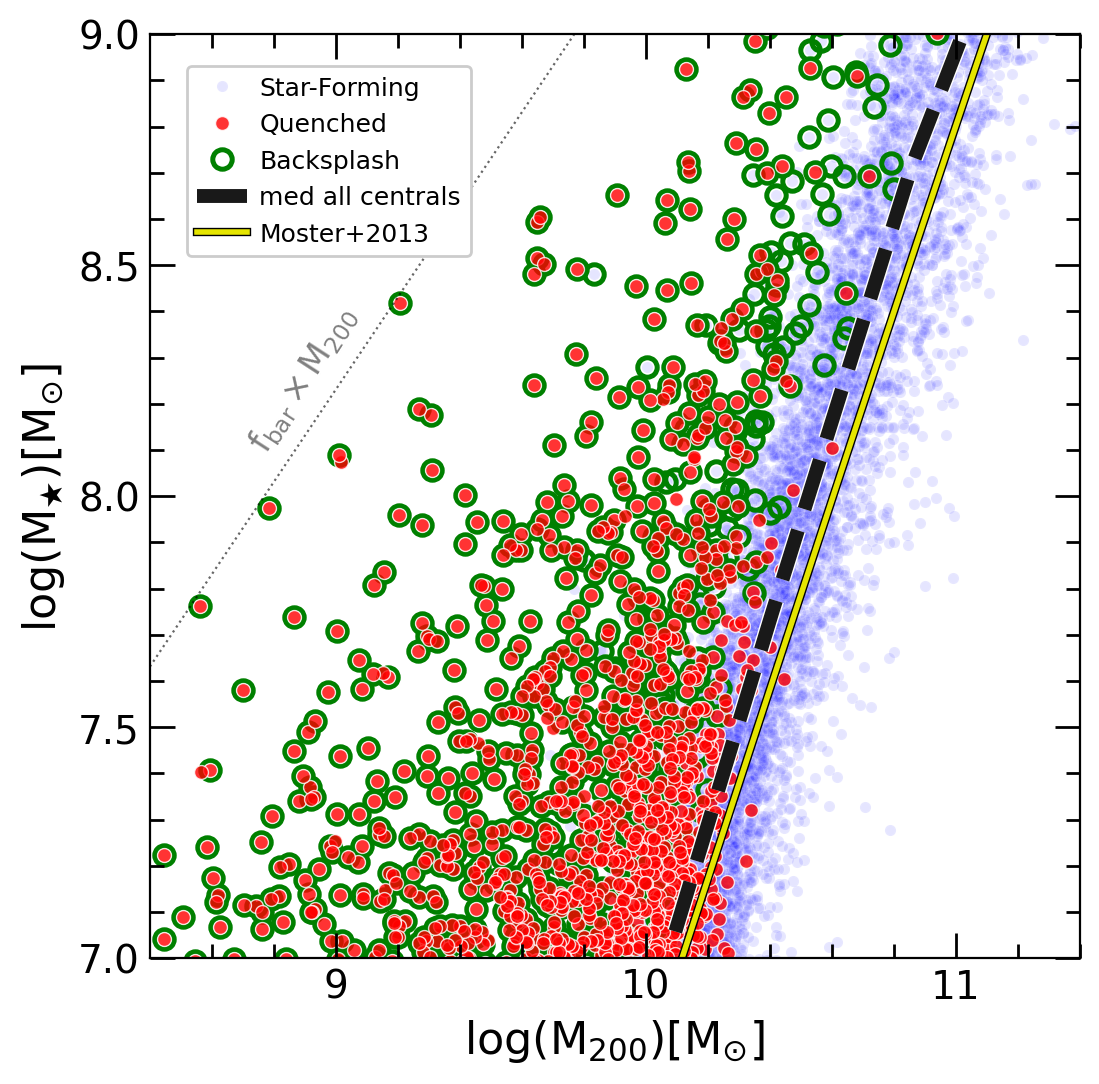}
	\caption{Halo mass–stellar mass relation ($M_{200}$-$M_{\star}$) for all central galaxies in TNG50, separated into star-forming (blue filled circles) and quenched (red filled circles) galaxies, with backsplash galaxies highlighted as green open circles (as in Fig.~\ref{fig:SSFR}). The black thick dashed line is the median of all central galaxies in TNG50, while yellow solid line represents the \citet{Moster2013} abundance-matching relation. Backsplash galaxies tend to lose a significant amount of dark matter, while quenched non-backsplash galaxies remain closer to the median $M_{200}$-$M_{\star}$ relation.}	\label{fig:MhaloMstar}
\end{figure}

We see this difference clearly in Fig.~\ref{fig:MhaloMstar}, where backsplash systems (green open symbols) are clearly offset from the main $M_{\star}$-$M_{200}$ trend outlined by most star-forming dwarfs (blue symbols). At given stellar mass, backsplash objects have in general lower halo masses, in some cases by more than one dex. On the other hand, quenched non-backsplash objects (red symbols without green circles) tend to outline the stellar mass - halo mass relation of the normal star-forming dwarf population (blue), highlighting the different impact of cosmic web stripping on halo mass compared to backsplash orbits.\\

\begin{figure*}
	\centering
	\includegraphics[width=\columnwidth]{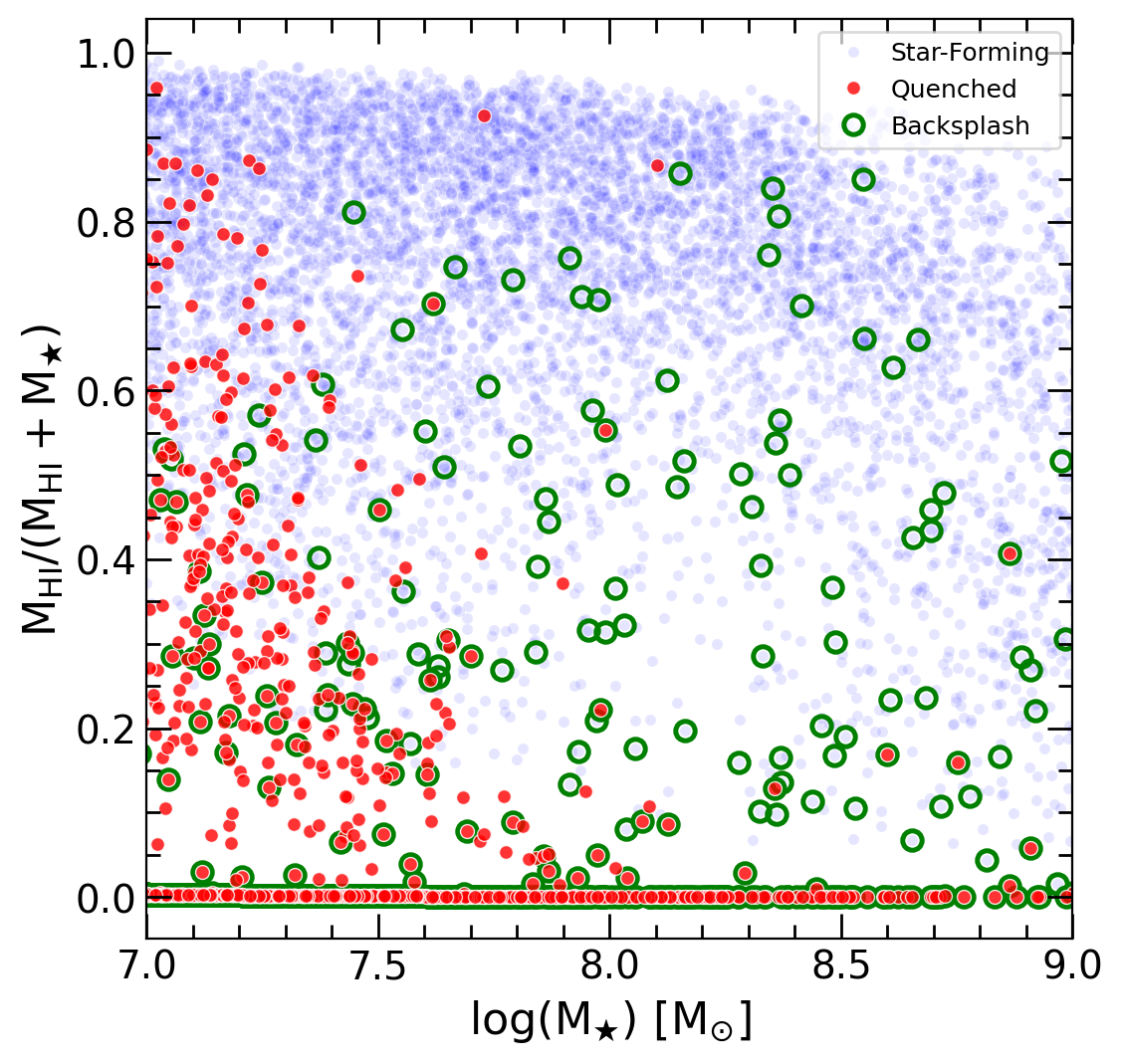}
	\includegraphics[width=\columnwidth]{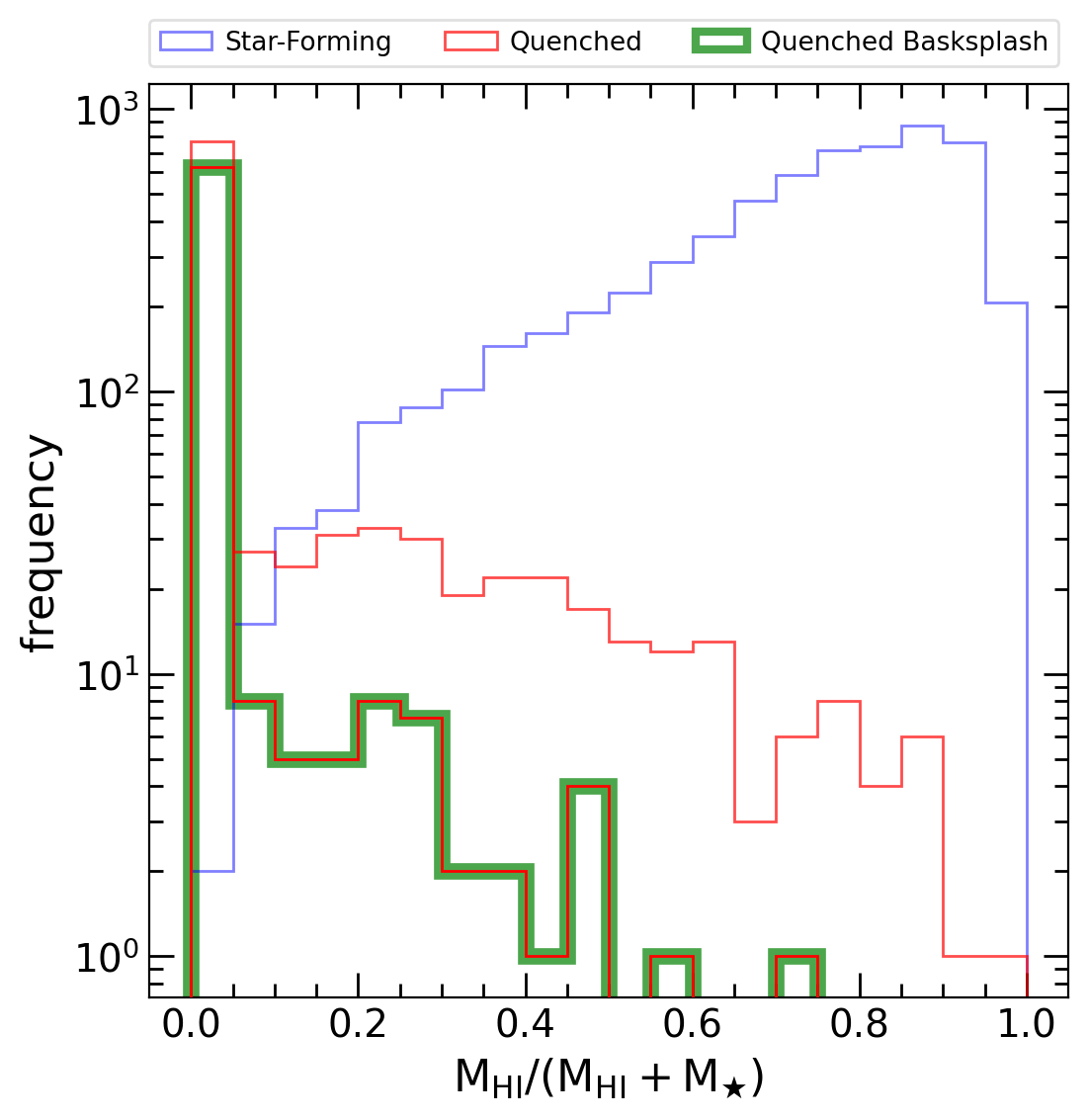}
	\caption{Fraction of neutral hydrogen ($HI$) mass with respect to baryon mass (neutral Hydrogen mass + stellar mass) as a function of stellar mass for all simulated central galaxies in TNG50 (left panel), separated into star-forming (blue filled circles) and quenched (red filled circles) galaxies, with backsplash galaxies highlighted as green open circles (as in previous figures). The right panel shows the distribution of the neutral Hydrogen fraction, with the distribution of backsplash quenched galaxies highlighted by thick green line, as in the left panel}.
	\label{fig:Mgas}
\end{figure*}

The gas content of quenched field dwarfs is shown in Fig.~\ref{fig:Mgas}. Typical star forming dwarfs in this mass range (blue symbols) have $\sim 80\%$ of their baryonic mass in the form of HI. Quenched objects (red) show a diverse gas content: while they are systematically below the star-forming population, some objects show gas fractions consistent with the lower tail of star-forming dwarfs while others are fully devoid of gas. This is better quantified in the distributions shown on the right panel of Fig.~\ref{fig:Mgas}. The median HI fraction in quenched field dwarfs (red thin line) is $0.28$ compared to the $0.77$ shown in the star-forming sample (blue thin line). 
Note that the gas depletion in backsplash objects (green thick line around thin red) tends to be more significant than in the rest of quenched dwarfs affected by cosmic web stripping. \\

\begin{figure*}
	\centering
	\includegraphics[width=\columnwidth]{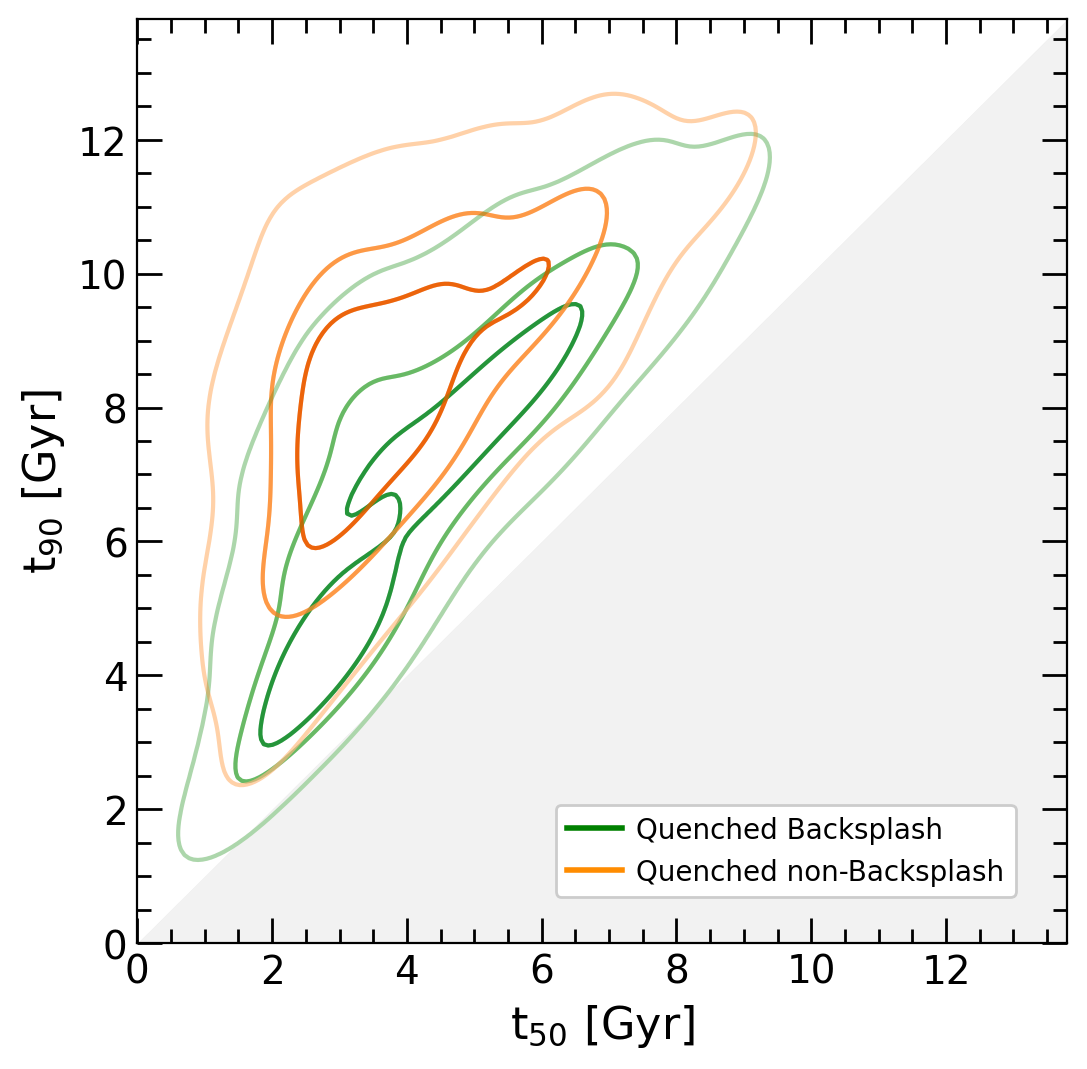}
	\includegraphics[width=\columnwidth]{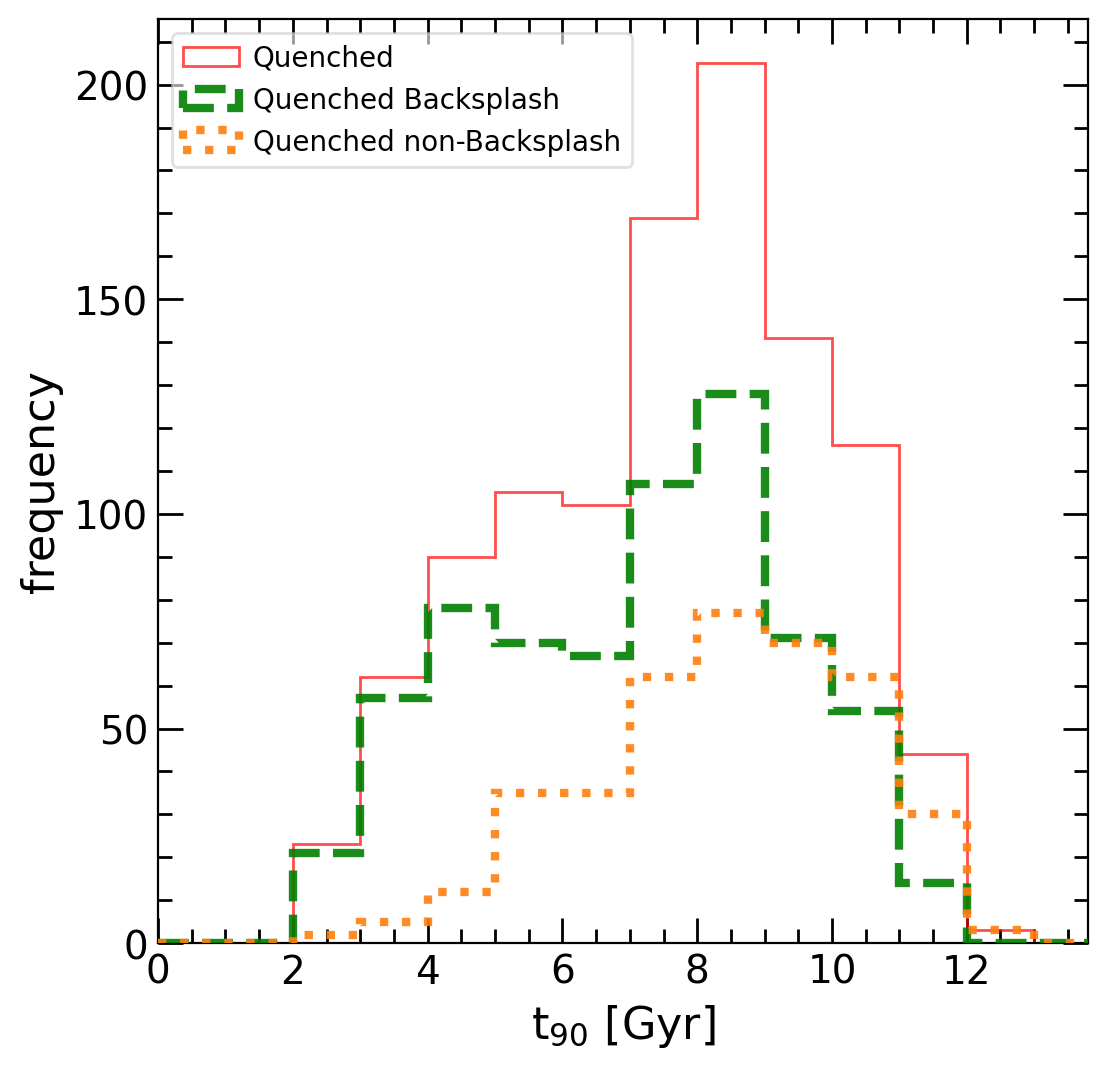}
	\caption{Left panel: Relation between the times to assemble 50\% ($t_{50}$) and 90\% ($t_{90}$) of stellar mass in cosmic time units ($z=0$ corresponds to $\sim 13.8$ Gyr). Quenched dwarf galaxies in TNG50 are separated into backsplash (green) and non-backsplash (orange) populations. Right panel: distribution of $t_{90}$ for all quenched dwarf galaxies in TNG50 (red histogram), separated into backsplash (green) and non-backsplash (orange) distributions. While the $t_{50}$ distribution is very similar for both populations of quenched dwarf galaxies, the $t_{90}$ values (indicative of quenching time) differ significantly, with non-backsplash galaxies exhibiting a later quenching timescale.}
	\label{fig:t50t90}
\end{figure*}
	%
\subsection{Backsplash or non-backsplash?}
\label{ssec:dwarfs_observations}

How would one decide observationally whether a particular field dwarf has been quenched as a result of backsplash effects, or of the cosmic web stripping discussed in the previous subsection? Backsplash leads to a reduction of the halo mass, which does not occur in the other cases, but measuring the total mass of a dwarf galaxy halo is exceedingly difficult. Empirically, we find that the time of quenching provides a relatively straightforward way to discriminate between quenching mechanisms.\\

We illustrate this in Figure~\ref{fig:t50t90}, where we show  when a particular galaxy has formed $50\%$ and $90\%$ of its final stellar mass ($t_{50}$ and $t_{90}$, respectively). The first timescale corresponds roughly to the median age of stars in a galaxy, and the latter is a crude but simple approximation of the quenching time (i.e., the time when a given galaxy stopped forming stars).\\

\begin{figure}
	\centering
	\includegraphics[width=\columnwidth]{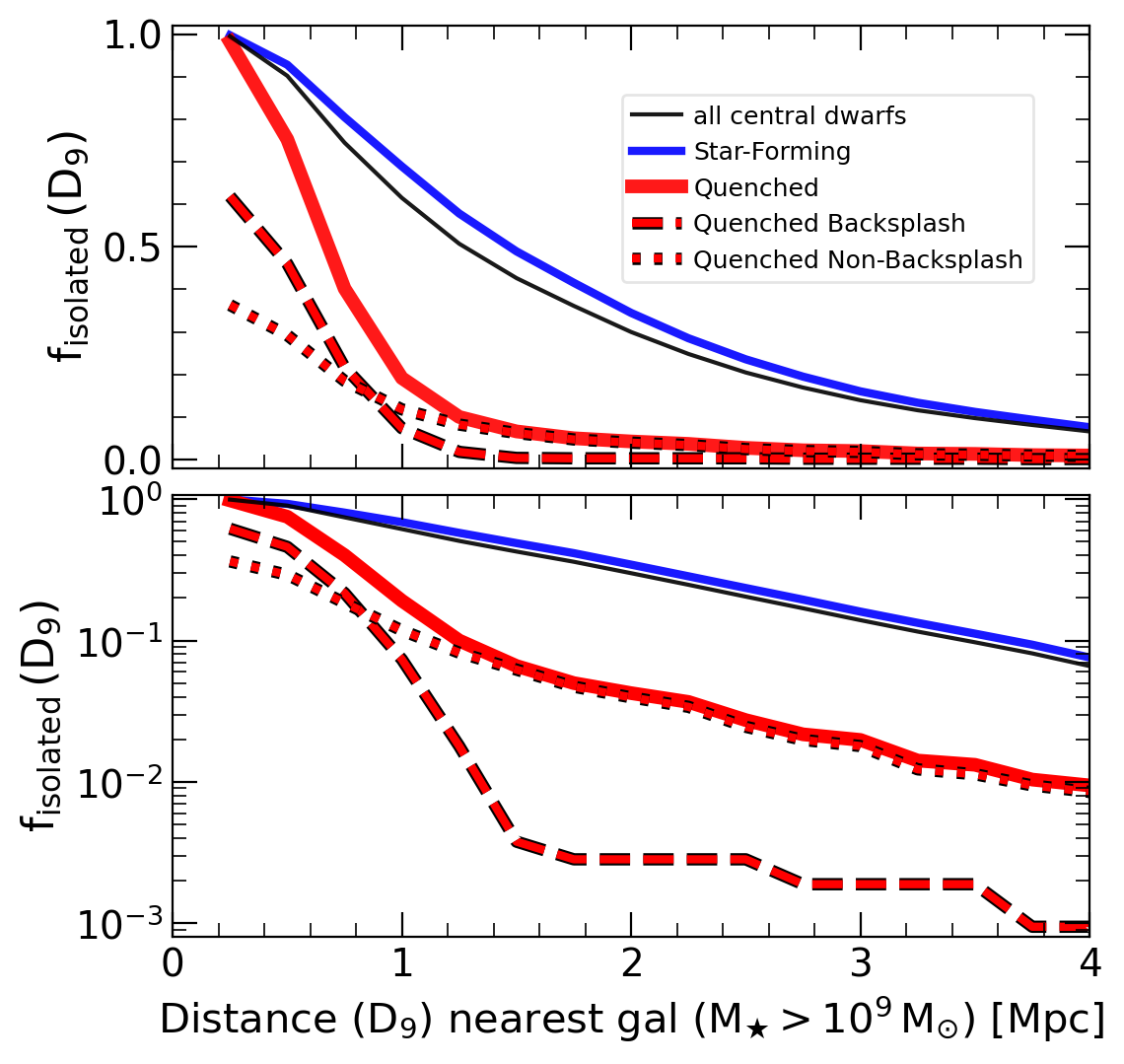}
	\caption{Fraction of dwarfs whose closest massive ($M_{\star}>10^9M_\odot$) neighbor lies beyond a distance $D_9$. The black thin line corresponds to all  central dwarf galaxies ($M_{\star} = [10^7 $-$ 10^9] ~ M_{\odot}$) in TNG50, while star-forming and quenched galaxies are highlighted with thick solid blue and red lines, respectively. In the case of quenched dwarf galaxies, we also indicate the contribution of  backsplash (dashed red line) and non-backsplash (dotted red line) galaxies to the quenched fraction (solid red).} 
	\label{fig:isolated_criteria}
\end{figure}

Field dwarf galaxies in TNG50 typically have $\bar{t}_{50}\sim 6.5\pm2.7$ Gyr ago and $\bar{t}_{90}< 1$ Gyr ago (i.e., most are still actively forming stars). As expected, quenched systems have, on average, older stellar populations with $\bar{t}_{50} \sim 10\pm1.8$ Gyr, and $\bar{t}_{90} \sim 6.5\pm2.5$ Gyr.
When restricting the sample to quenched dwarfs that are backsplash, the median $t_{50}$ and $t_{90}$ are $4.0^{+1.7}_{-1.2}$ and $7.4^{+1.4}_{-2.1}$ Gyr, respectively; while, for non-backsplash dwarf galaxies these values are $3.7^{+1.4}_{-0.9}$ and $8.5^{+1.4}_{-1.3}$ Gyr. (The uncertainties in the numbers quoted above indicate the $25$-$75$ percentile range.) Therefore, dwarfs that have been quenched by cosmic web stripping have, on average, similar $t_{50}$ as their backsplash counterparts but they show a slight but systematic shift towards later times for $t_{90}$.
The preference for later quenching times in cosmic web stripped dwarfs is also evident from the right panel in Fig.~\ref{fig:t50t90} showing the $t_{90}$ distributions.\\

The interpretation seems clear. For a backsplash galaxy to be in the field at present, the interaction with the massive host must have happened sufficiently early to allow the dwarf to reach a low-density location. This favors earlier quenching times for backsplash dwarfs compared with systems that have been quenched by cosmic web stripping. Quenching in these cases can happen more gradually, favoring later quenching times The overall effect is, however, small, and unlikely to provide a clean test to discriminate between backsplash and non-backsplash systems.\\

An additional criteria that can be used to help assess the origin of quenched dwarfs is their degree of isolation. Fig.~\ref{fig:isolated_criteria} shows the fraction of dwarfs whose closest massive neighbour (defined as a system with $M_{\star} = 10^9~ M_{\odot}$ or above) is beyond a given distance $D_9$. A solid black line indicates the results for all field dwarfs, showing that $90$\% of them have their closest massive galaxy within $\sim 500$ kpc. One half of them have the closest massive companion located within $1.25$ Mpc. By contrast, only $\lesssim 7\%$ of dwarfs are so isolated that the closest massive companion is at $>4$ Mpc.\\

Most field dwarfs are star-forming, and therefore the results for all field dwarfs (solid black) and star-forming dwarfs (solid blue) are similar. On the other hand, quenched systems (solid red) behave differently. Most quenched dwarfs have nearby massive companions, as expected given the importance of the backsplash mechanism discussed earlier.  Half of all quenched field dwarfs have a massive neighbour within $\sim 700$ kpc, and only $10\%$ of them have its closest massive neighbor beyond $\sim 1.25$ Mpc. Adopting  stricter isolation criteria reduces these fractions even further. Only about$1\%$ of quenched dwarfs have their closest companion at a distance greater than 4 Mpc, compared to $\sim 8\%$ for star-forming dwarfs.\\

The dashed-red and dotted-red lines in Fig.~\ref{fig:isolated_criteria} split the quenched dwarf sample into backsplash objects and cosmic web stripped systems, respectively.  This shows that backsplash becomes subdominant for $D_9 > 1$ Mpc and that its importance decreases further with increasing $D_9$. For quenched systems without a massive neighbour closer than $1.5$ Mpc, only $\sim 6\%$ are backsplash. This implies that, in practice, deeply isolated quenched dwarfs (i.e., $D_9 > 1.5$ Mpc) are in all likelihood the result of cosmic web stripping. However, only $\sim 1\%$ of deeply isolated dwarfs are quenched, so searching for this smoking-gun evidence of the effects of cosmic web stripping will prove challenging.\\

\section{Summary and Conclusions}
\label{SecConc}

We have used the cosmological hydrodynamical simulation TNG50 to study the quenched fraction of dwarf galaxies as a function of galaxy mass and environment, with particular emphasis on the mechanisms that lead to the quenching of star formation in field dwarfs. This is an important issue because it is fairly well established observationally  that most, if not all, quenched dwarfs are satellites or former satellites of more massive hosts, and that few, if any, quenched dwarf galaxies exist in the field.\\

Our analysis of TNG50 shows that quenched field (i.e., not satellites) dwarfs are relatively rare, reaching on average $15\%$ of the population in the $M_\star=10^7 - 10^9 ~ M_{\odot}$ range studied here. Most field dwarfs are quenched by backsplash, and, therefore, lie typically in close proximity to a massive host.  Restricting the analysis to deeply isolated systems, defined as those whose closest neighbour exceeding $M_{\star}=10^9\, M_\odot$ is farther than $D_9 = 1.5$ Mpc, reduces further the quenched fraction to $\sim1\%$, in general agreement with observational constraints.\\

In the simulation, the field dwarf quenched fraction increases towards lower masses, from a minimum of $2.2\%$ at $M_{\star}\sim 10^9\, M_\odot$. This increase is seen even after excluding all backsplash systems, which hints at the presence of an additional mechanism responsible for the cessation of star formation in low-mass systems.\\

We show here that non-backsplash isolated dwarfs quench because of environmental effects, namely ram-pressure stripping by the cosmic web, as field dwarfs join or pass through filaments or other overdense regions. These systems may, as a group, be distinguished from backsplash dwarfs because of their relatively late quenching times and because of their isolation. Of all quenched dwarfs with $D_9>1.5$ Mpc, only $\sim 6\%$ are backsplash objects. The latter also quench earlier: their median quenching time ($t_{90}$) is $7.4$ Gyr, compared with $8.5$ Gyr for those affected by cosmic web stripping.\\

The efficiency of cosmic web stripping is expected to increase towards lower masses, and may very well be an important quenching mechanism in very faint systems, whose quenched fractions and timescales are not yet well known, either in simulations or in observations.\\

The main conclusion of our analysis is that some quenched dwarf galaxies are expected to exist in the field, and that their numbers should increase towards fainter systems. Detecting such systems in samples of deeply isolated dwarf galaxies, or in regions of extremely low galaxy density, such as cosmic voids, would provide support for this interpretation, and for the idea that the cosmic web may actually be responsible for shaping the evolutionary history of the faintest galaxies.

\section*{Acknowledgments}
JAB and LVS is grateful for partial financial support from NSF-CAREER-1945310 and NSF-AST-2107993 grants. Computations were performed using the computer clusters and data storage resources of the HPCC, which were funded by grants from NSF (MRI-2215705, MRI-1429826) and NIH (1S10OD016290-01A1). JFN acknowledges the hospitality of the Max-Planck Institute for Astrophysics and of the Donostia International Physics Center during the writing of this manuscript.

\section*{Data Availability}

This paper is based on halo catalogs and merger trees from the Illustris-TNG Project \citep{Nelson2019TNG, Nelson2019TNG50}. These data are publicly available at \href{https://www.tng-project.org/}{https://www.tng-project.org/}. The main properties of the dwarf galaxy samples, and other products included in this analysis, may be shared upon request to the corresponding author if no further conflict exists with ongoing projects. 


\bibliographystyle{aasjournal}
\bibliography{biblio}

\begin{thebibliography}{}
\expandafter\ifx\csname natexlab\endcsname\relax\def\natexlab#1{#1}\fi
\providecommand{\url}[1]{\href{#1}{#1}}
\providecommand{\dodoi}[1]{doi:~\href{http://doi.org/#1}{\nolinkurl{#1}}}
\providecommand{\doeprint}[1]{\href{http://ascl.net/#1}{\nolinkurl{http://ascl.net/#1}}}
\providecommand{\doarXiv}[1]{\href{https://arxiv.org/abs/#1}{\nolinkurl{https://arxiv.org/abs/#1}}}

\bibitem[{{Abadi} {et~al.}(1999){Abadi}, {Moore}, \& {Bower}}]{Abadi1999}
{Abadi}, M.~G., {Moore}, B., \& {Bower}, R.~G. 1999, \mnras, 308, 947,
  \dodoi{10.1046/j.1365-8711.1999.02715.x}

\bibitem[{{Balogh} {et~al.}(2000){Balogh}, {Navarro}, \& {Morris}}]{Balogh2000}
{Balogh}, M.~L., {Navarro}, J.~F., \& {Morris}, S.~L. 2000, \apj, 540, 113,
  \dodoi{10.1086/309323}

\bibitem[{{Benavides} {et~al.}(2021){Benavides}, {Sales}, {Abadi}, {Pillepich},
  {Nelson}, {Marinacci}, {Cooper}, {Pakmor}, {Torrey}, {Vogelsberger}, \&
  {Hernquist}}]{Benavides2021}
{Benavides}, J.~A., {Sales}, L.~V., {Abadi}, M.~G., {et~al.} 2021, Nature
  Astronomy, 5, 1255, \dodoi{10.1038/s41550-021-01458-1}

\bibitem[{{Benitez-Llambay} \& {Frenk}(2020)}]{Benitez-Llambay2020}
{Benitez-Llambay}, A., \& {Frenk}, C. 2020, \mnras, 498, 4887,
  \dodoi{10.1093/mnras/staa2698}

\bibitem[{{Ben{\'\i}tez-Llambay} {et~al.}(2013){Ben{\'\i}tez-Llambay},
  {Navarro}, {Abadi}, {Gottl{\"o}ber}, {Yepes}, {Hoffman}, \&
  {Steinmetz}}]{Benitez-Llambay2013}
{Ben{\'\i}tez-Llambay}, A., {Navarro}, J.~F., {Abadi}, M.~G., {et~al.} 2013,
  \apjl, 763, L41, \dodoi{10.1088/2041-8205/763/2/L41}

\bibitem[{{Ben{\'\i}tez-Llambay} {et~al.}(2017){Ben{\'\i}tez-Llambay},
  {Navarro}, {Frenk}, {Sawala}, {Oman}, {Fattahi}, {Schaller}, {Schaye},
  {Crain}, \& {Theuns}}]{Benitez-Llambay2017}
{Ben{\'\i}tez-Llambay}, A., {Navarro}, J.~F., {Frenk}, C.~S., {et~al.} 2017,
  \mnras, 465, 3913, \dodoi{10.1093/mnras/stw2982}

\bibitem[{{Bhattacharyya} {et~al.}(2025){Bhattacharyya}, {Peter}, \&
  {Leauthaud}}]{Bhattacharyya2025}
{Bhattacharyya}, J., {Peter}, A. H.~G., \& {Leauthaud}, A. 2025, arXiv
  e-prints, arXiv:2501.01946, \dodoi{10.48550/arXiv.2501.01946}

\bibitem[{{Bidaran} {et~al.}(2025){Bidaran}, {P{\'e}rez},
  {S{\'a}nchez-Menguiano}, {Argudo-Fern{\'a}ndez}, {Ferr{\'e}-Mateu},
  {Navarro}, {Peletier}, {Ruiz-Lara}, {van de Ven}, {Verley}, {Zurita}, {Duarte
  Puertas}, {Falc{\'o}n-Barroso}, {S{\'a}nchez-Bl{\'a}zquez}, \&
  {Jim{\'e}nez}}]{Bidaran2025}
{Bidaran}, B., {P{\'e}rez}, I., {S{\'a}nchez-Menguiano}, L., {et~al.} 2025,
  arXiv e-prints, arXiv:2501.02910, \dodoi{10.48550/arXiv.2501.02910}

\bibitem[{{Boylan-Kolchin} {et~al.}(2011){Boylan-Kolchin}, {Bullock}, \&
  {Kaplinghat}}]{Boylan-Kolchin2011}
{Boylan-Kolchin}, M., {Bullock}, J.~S., \& {Kaplinghat}, M. 2011, \mnras, 415,
  L40, \dodoi{10.1111/j.1745-3933.2011.01074.x}

\bibitem[{{Busha} {et~al.}(2010){Busha}, {Alvarez}, {Wechsler}, {Abel}, \&
  {Strigari}}]{Busha2010}
{Busha}, M.~T., {Alvarez}, M.~A., {Wechsler}, R.~H., {Abel}, T., \& {Strigari},
  L.~E. 2010, \apj, 710, 408, \dodoi{10.1088/0004-637X/710/1/408}

\bibitem[{{Carleton} {et~al.}(2024){Carleton}, {Ellsworth-Bowers}, {Windhorst},
  {Cohen}, {Conselice}, {Diego}, {Zitrin}, {Archer}, {McIntyre}, {Kamieneski},
  {Jansen}, {Summers}, {D'Silva}, {Koekemoer}, {Coe}, {Driver}, {Frye},
  {Grogin}, {Marshall}, {Nonino}, {Pirzkal}, {Robotham}, {Ryan}, {Ortiz},
  {Tompkins}, {Willmer}, {Yan}, \& {Holwerda}}]{Carleton2024}
{Carleton}, T., {Ellsworth-Bowers}, T., {Windhorst}, R.~A., {et~al.} 2024,
  \apjl, 961, L37, \dodoi{10.3847/2041-8213/ad1b56}

\bibitem[{{Casey} {et~al.}(2023){Casey}, {Greco}, {Peter}, \&
  {Davis}}]{Casey2023}
{Casey}, K.~J., {Greco}, J.~P., {Peter}, A. H.~G., \& {Davis}, A.~B. 2023,
  \mnras, 520, 4715, \dodoi{10.1093/mnras/stad352}

\bibitem[{{Chan} {et~al.}(2018){Chan}, {Kere{\v{s}}}, {Wetzel}, {Hopkins},
  {Faucher-Gigu{\`e}re}, {El-Badry}, {Garrison-Kimmel}, \&
  {Boylan-Kolchin}}]{Chan2018}
{Chan}, T.~K., {Kere{\v{s}}}, D., {Wetzel}, A., {et~al.} 2018, \mnras, 478,
  906, \dodoi{10.1093/mnras/sty1153}

\bibitem[{{Davis} {et~al.}(1985){Davis}, {Efstathiou}, {Frenk}, \&
  {White}}]{Davis1985}
{Davis}, M., {Efstathiou}, G., {Frenk}, C.~S., \& {White}, S.~D.~M. 1985, \apj,
  292, 371, \dodoi{10.1086/163168}

\bibitem[{{Dolag} {et~al.}(2009){Dolag}, {Borgani}, {Murante}, \&
  {Springel}}]{Dolag2009}
{Dolag}, K., {Borgani}, S., {Murante}, G., \& {Springel}, V. 2009, \mnras, 399,
  497, \dodoi{10.1111/j.1365-2966.2009.15034.x}

\bibitem[{{Geha} {et~al.}(2012){Geha}, {Blanton}, {Yan}, \&
  {Tinker}}]{Geha2012}
{Geha}, M., {Blanton}, M.~R., {Yan}, R., \& {Tinker}, J.~L. 2012, \apj, 757,
  85, \dodoi{10.1088/0004-637X/757/1/85}

\bibitem[{{Gunn} \& {Gott}(1972)}]{GunnGott1972}
{Gunn}, J.~E., \& {Gott}, J.~Richard, I. 1972, \apj, 176, 1,
  \dodoi{10.1086/151605}

\bibitem[{{Herzog} {et~al.}(2023){Herzog}, {Ben{\'\i}tez-Llambay}, \&
  {Fumagalli}}]{Herzog2023}
{Herzog}, G., {Ben{\'\i}tez-Llambay}, A., \& {Fumagalli}, M. 2023, \mnras, 518,
  6305, \dodoi{10.1093/mnras/stac3282}

\bibitem[{{Li} {et~al.}(2024){Li}, {Greene}, {Carlsten}, \& {Danieli}}]{Li2024}
{Li}, J., {Greene}, J.~E., {Carlsten}, S.~G., \& {Danieli}, S. 2024, \apjl,
  975, L23, \dodoi{10.3847/2041-8213/ad5b59}

\bibitem[{{Ludlow} {et~al.}(2009){Ludlow}, {Navarro}, {Springel}, {Jenkins},
  {Frenk}, \& {Helmi}}]{Ludlow2009ApJ}
{Ludlow}, A.~D., {Navarro}, J.~F., {Springel}, V., {et~al.} 2009, \apj, 692,
  931, \dodoi{10.1088/0004-637X/692/1/931}

\bibitem[{{Mamon} {et~al.}(2004){Mamon}, {Sanchis}, {Salvador-Sol{\'e}}, \&
  {Solanes}}]{Mamon2004}
{Mamon}, G.~A., {Sanchis}, T., {Salvador-Sol{\'e}}, E., \& {Solanes}, J.~M.
  2004, \aap, 414, 445, \dodoi{10.1051/0004-6361:20034155}

\bibitem[{{Moster} {et~al.}(2013){Moster}, {Naab}, \& {White}}]{Moster2013}
{Moster}, B.~P., {Naab}, T., \& {White}, S. D.~M. 2013, \mnras, 428, 3121,
  \dodoi{10.1093/mnras/sts261}

\bibitem[{{Muriel} \& {Coenda}(2014)}]{Muriel2014}
{Muriel}, H., \& {Coenda}, V. 2014, \aap, 564, A85,
  \dodoi{10.1051/0004-6361/201322033}

\bibitem[{{Nelson} {et~al.}(2018){Nelson}, {Pillepich}, {Springel},
  {Weinberger}, {Hernquist}, {Pakmor}, {Genel}, {Torrey}, {Vogelsberger},
  {Kauffmann}, {Marinacci}, \& {Naiman}}]{Nelson2018}
{Nelson}, D., {Pillepich}, A., {Springel}, V., {et~al.} 2018, \mnras, 475, 624,
  \dodoi{10.1093/mnras/stx3040}

\bibitem[{{Nelson} {et~al.}(2019{\natexlab{a}}){Nelson}, {Springel},
  {Pillepich}, {Rodriguez-Gomez}, {Torrey}, {Genel}, {Vogelsberger}, {Pakmor},
  {Marinacci}, {Weinberger}, {Kelley}, {Lovell}, {Diemer}, \&
  {Hernquist}}]{Nelson2019TNG}
{Nelson}, D., {Springel}, V., {Pillepich}, A., {et~al.} 2019{\natexlab{a}},
  Computational Astrophysics and Cosmology, 6, 2,
  \dodoi{10.1186/s40668-019-0028-x}

\bibitem[{{Nelson} {et~al.}(2019{\natexlab{b}}){Nelson}, {Pillepich},
  {Springel}, {Pakmor}, {Weinberger}, {Genel}, {Torrey}, {Vogelsberger},
  {Marinacci}, \& {Hernquist}}]{Nelson2019TNG50}
{Nelson}, D., {Pillepich}, A., {Springel}, V., {et~al.} 2019{\natexlab{b}},
  \mnras, 490, 3234, \dodoi{10.1093/mnras/stz2306}

\bibitem[{{Okamoto} \& {Frenk}(2009)}]{Okamoto2009}
{Okamoto}, T., \& {Frenk}, C.~S. 2009, \mnras, 399, L174,
  \dodoi{10.1111/j.1745-3933.2009.00748.x}

\bibitem[{{Park} {et~al.}(2023){Park}, {Belli}, {Conroy}, {Tacchella}, {Leja},
  {Cutler}, {Johnson}, {Nelson}, \& {Emami}}]{Park2023}
{Park}, M., {Belli}, S., {Conroy}, C., {et~al.} 2023, \apj, 953, 119,
  \dodoi{10.3847/1538-4357/acd54a}

\bibitem[{{Pasha} {et~al.}(2023){Pasha}, {Mandelker}, {van den Bosch},
  {Springel}, \& {van de Voort}}]{Pasha2023}
{Pasha}, I., {Mandelker}, N., {van den Bosch}, F.~C., {Springel}, V., \& {van
  de Voort}, F. 2023, \mnras, 520, 2692, \dodoi{10.1093/mnras/stac3776}

\bibitem[{{Pereira-Wilson} {et~al.}(2023){Pereira-Wilson}, {Navarro},
  {Ben{\'\i}tez-Llambay}, \& {Santos-Santos}}]{Pereira-Wilson2023}
{Pereira-Wilson}, M., {Navarro}, J.~F., {Ben{\'\i}tez-Llambay}, A., \&
  {Santos-Santos}, I. 2023, \mnras, 519, 1425, \dodoi{10.1093/mnras/stac3633}

\bibitem[{{Pillepich} {et~al.}(2018{\natexlab{a}}){Pillepich}, {Springel},
  {Nelson}, {Genel}, {Naiman}, {Pakmor}, {Hernquist}, {Torrey}, {Vogelsberger},
  {Weinberger}, \& {Marinacci}}]{Pillepich2018a}
{Pillepich}, A., {Springel}, V., {Nelson}, D., {et~al.} 2018{\natexlab{a}},
  \mnras, 473, 4077, \dodoi{10.1093/mnras/stx2656}

\bibitem[{{Pillepich} {et~al.}(2018{\natexlab{b}}){Pillepich}, {Nelson},
  {Hernquist}, {Springel}, {Pakmor}, {Torrey}, {Weinberger}, {Genel}, {Naiman},
  {Marinacci}, \& {Vogelsberger}}]{Pillepich2018b}
{Pillepich}, A., {Nelson}, D., {Hernquist}, L., {et~al.} 2018{\natexlab{b}},
  \mnras, 475, 648, \dodoi{10.1093/mnras/stx3112}

\bibitem[{{Pillepich} {et~al.}(2019){Pillepich}, {Nelson}, {Springel},
  {Pakmor}, {Torrey}, {Weinberger}, {Vogelsberger}, {Marinacci}, {Genel}, {van
  der Wel}, \& {Hernquist}}]{Pillepich2019}
{Pillepich}, A., {Nelson}, D., {Springel}, V., {et~al.} 2019, \mnras, 490,
  3196, \dodoi{10.1093/mnras/stz2338}

\bibitem[{{Planck Collaboration} {et~al.}(2016){Planck Collaboration}, {Ade},
  {Aghanim}, {Arnaud}, {Ashdown}, {Aumont}, {Baccigalupi}, {Banday},
  {Barreiro}, {Bartlett}, {Bartolo}, {Battaner}, {Battye}, {Benabed},
  {Beno{\^\i}t}, {Benoit-L{\'e}vy}, {Bernard}, {Bersanelli}, {Bielewicz},
  {Bock}, {Bonaldi}, {Bonavera}, {Bond}, {Borrill}, {Bouchet}, {Boulanger},
  {Bucher}, {Burigana}, {Butler}, {Calabrese}, {Cardoso}, {Catalano},
  {Challinor}, {Chamballu}, {Chary}, {Chiang}, {Chluba}, {Christensen},
  {Church}, {Clements}, {Colombi}, {Colombo}, {Combet}, {Coulais}, {Crill},
  {Curto}, {Cuttaia}, {Danese}, {Davies}, {Davis}, {de Bernardis}, {de Rosa},
  {de Zotti}, {Delabrouille}, {D{\'e}sert}, {Di Valentino}, {Dickinson},
  {Diego}, {Dolag}, {Dole}, {Donzelli}, {Dor{\'e}}, {Douspis}, {Ducout},
  {Dunkley}, {Dupac}, {Efstathiou}, {Elsner}, {En{\ss}lin}, {Eriksen},
  {Farhang}, {Fergusson}, {Finelli}, {Forni}, {Frailis}, {Fraisse},
  {Franceschi}, {Frejsel}, {Galeotta}, {Galli}, {Ganga}, {Gauthier}, {Gerbino},
  {Ghosh}, {Giard}, {Giraud-H{\'e}raud}, {Giusarma}, {Gjerl{\o}w},
  {Gonz{\'a}lez-Nuevo}, {G{\'o}rski}, {Gratton}, {Gregorio}, {Gruppuso},
  {Gudmundsson}, {Hamann}, {Hansen}, {Hanson}, {Harrison}, {Helou},
  {Henrot-Versill{\'e}}, {Hern{\'a}ndez-Monteagudo}, {Herranz}, {Hildebrandt},
  {Hivon}, {Hobson}, {Holmes}, {Hornstrup}, {Hovest}, {Huang}, {Huffenberger},
  {Hurier}, {Jaffe}, {Jaffe}, {Jones}, {Juvela}, {Keih{\"a}nen}, {Keskitalo},
  {Kisner}, {Kneissl}, {Knoche}, {Knox}, {Kunz}, {Kurki-Suonio}, {Lagache},
  {L{\"a}hteenm{\"a}ki}, {Lamarre}, {Lasenby}, {Lattanzi}, {Lawrence}, {Leahy},
  {Leonardi}, {Lesgourgues}, {Levrier}, {Lewis}, {Liguori}, {Lilje},
  {Linden-V{\o}rnle}, {L{\'o}pez-Caniego}, {Lubin}, {Mac{\'\i}as-P{\'e}rez},
  {Maggio}, {Maino}, {Mandolesi}, {Mangilli}, {Marchini}, {Maris}, {Martin},
  {Martinelli}, {Mart{\'\i}nez-Gonz{\'a}lez}, {Masi}, {Matarrese}, {McGehee},
  {Meinhold}, {Melchiorri}, {Melin}, {Mendes}, {Mennella}, {Migliaccio},
  {Millea}, {Mitra}, {Miville-Desch{\^e}nes}, {Moneti}, {Montier}, {Morgante},
  {Mortlock}, {Moss}, {Munshi}, {Murphy}, {Naselsky}, {Nati}, {Natoli},
  {Netterfield}, {N{\o}rgaard-Nielsen}, {Noviello}, {Novikov}, {Novikov},
  {Oxborrow}, {Paci}, {Pagano}, {Pajot}, {Paladini}, {Paoletti}, {Partridge},
  {Pasian}, {Patanchon}, {Pearson}, {Perdereau}, {Perotto}, {Perrotta},
  {Pettorino}, {Piacentini}, {Piat}, {Pierpaoli}, {Pietrobon}, {Plaszczynski},
  {Pointecouteau}, {Polenta}, {Popa}, {Pratt}, {Pr{\'e}zeau}, {Prunet},
  {Puget}, {Rachen}, {Reach}, {Rebolo}, {Reinecke}, {Remazeilles}, {Renault},
  {Renzi}, {Ristorcelli}, {Rocha}, {Rosset}, {Rossetti}, {Roudier},
  {Rouill{\'e} d'Orfeuil}, {Rowan-Robinson}, {Rubi{\~n}o-Mart{\'\i}n},
  {Rusholme}, {Said}, {Salvatelli}, {Salvati}, {Sandri}, {Santos},
  {Savelainen}, {Savini}, {Scott}, {Seiffert}, {Serra}, {Shellard}, {Spencer},
  {Spinelli}, {Stolyarov}, {Stompor}, {Sudiwala}, {Sunyaev}, {Sutton},
  {Suur-Uski}, {Sygnet}, {Tauber}, {Terenzi}, {Toffolatti}, {Tomasi},
  {Tristram}, {Trombetti}, {Tucci}, {Tuovinen}, {T{\"u}rler}, {Umana},
  {Valenziano}, {Valiviita}, {Van Tent}, {Vielva}, {Villa}, {Wade}, {Wandelt},
  {Wehus}, {White}, {White}, {Wilkinson}, {Yvon}, {Zacchei}, \&
  {Zonca}}]{PlankColaboration2016}
{Planck Collaboration}, {Ade}, P.~A.~R., {Aghanim}, N., {et~al.} 2016, \aap,
  594, A13, \dodoi{10.1051/0004-6361/201525830}

\bibitem[{{Polzin} {et~al.}(2021){Polzin}, {van Dokkum}, {Danieli}, {Greco}, \&
  {Romanowsky}}]{Polzin2021}
{Polzin}, A., {van Dokkum}, P., {Danieli}, S., {Greco}, J.~P., \& {Romanowsky},
  A.~J. 2021, \apjl, 914, L23, \dodoi{10.3847/2041-8213/ac024f}

\bibitem[{{Prole} {et~al.}(2021){Prole}, {van der Burg}, {Hilker}, \&
  {Spitler}}]{Prole2021}
{Prole}, D.~J., {van der Burg}, R.~F.~J., {Hilker}, M., \& {Spitler}, L.~R.
  2021, \mnras, 500, 2049, \dodoi{10.1093/mnras/staa3296}

\bibitem[{{Rodriguez-Gomez} {et~al.}(2015){Rodriguez-Gomez}, {Genel},
  {Vogelsberger}, {Sijacki}, {Pillepich}, {Sales}, {Torrey}, {Snyder},
  {Nelson}, {Springel}, {Ma}, \& {Hernquist}}]{RodriguezGomez2015}
{Rodriguez-Gomez}, V., {Genel}, S., {Vogelsberger}, M., {et~al.} 2015, \mnras,
  449, 49, \dodoi{10.1093/mnras/stv264}

\bibitem[{{Rom{\'a}n} {et~al.}(2019){Rom{\'a}n}, {Beasley}, {Ruiz-Lara}, \&
  {Valls-Gabaud}}]{roman2019}
{Rom{\'a}n}, J., {Beasley}, M.~A., {Ruiz-Lara}, T., \& {Valls-Gabaud}, D. 2019,
  \mnras, 486, 823, \dodoi{10.1093/mnras/stz835}

\bibitem[{{Sales} {et~al.}(2007){Sales}, {Navarro}, {Abadi}, \&
  {Steinmetz}}]{Sales2007}
{Sales}, L.~V., {Navarro}, J.~F., {Abadi}, M.~G., \& {Steinmetz}, M. 2007,
  \mnras, 379, 1475, \dodoi{10.1111/j.1365-2966.2007.12026.x}

\bibitem[{{Samuel} {et~al.}(2023){Samuel}, {Pardasani}, {Wetzel},
  {Santistevan}, {Boylan-Kolchin}, {Moreno}, \&
  {Faucher-Gigu{\`e}re}}]{Samuel2023}
{Samuel}, J., {Pardasani}, B., {Wetzel}, A., {et~al.} 2023, \mnras, 525, 3849,
  \dodoi{10.1093/mnras/stad2576}

\bibitem[{{Sand} {et~al.}(2022){Sand}, {Mutlu-Pakdil}, {Jones}, {Karunakaran},
  {Wang}, {Yang}, {Chiti}, {Bennet}, {Crnojevi{\'c}}, \& {Spekkens}}]{Sand2022}
{Sand}, D.~J., {Mutlu-Pakdil}, B., {Jones}, M.~G., {et~al.} 2022, \apjl, 935,
  L17, \dodoi{10.3847/2041-8213/ac85ee}

\bibitem[{{Santos-Santos} {et~al.}(2023){Santos-Santos}, {Navarro}, \&
  {McConnachie}}]{Santos-Santos2023}
{Santos-Santos}, I. M.~E., {Navarro}, J.~F., \& {McConnachie}, A. 2023, \mnras,
  520, 55, \dodoi{10.1093/mnras/stad085}

\bibitem[{{Springel}(2010)}]{Springel2010}
{Springel}, V. 2010, \mnras, 401, 791, \dodoi{10.1111/j.1365-2966.2009.15715.x}

\bibitem[{{Springel} {et~al.}(2001){Springel}, {White}, {Tormen}, \&
  {Kauffmann}}]{Springel2001}
{Springel}, V., {White}, S. D.~M., {Tormen}, G., \& {Kauffmann}, G. 2001,
  \mnras, 328, 726, \dodoi{10.1046/j.1365-8711.2001.04912.x}

\bibitem[{{Stephenson} {et~al.}(2024){Stephenson}, {Stott}, {Butler},
  {Webster}, \& {Head}}]{Stephenson2024}
{Stephenson}, H., {Stott}, J., {Butler}, J., {Webster}, M., \& {Head}, J. 2024,
  arXiv e-prints, arXiv:2412.07834.
\newblock \doarXiv{2412.07834}

\bibitem[{{Teyssier} {et~al.}(2012){Teyssier}, {Johnston}, \&
  {Kuhlen}}]{Teyssier2012}
{Teyssier}, M., {Johnston}, K.~V., \& {Kuhlen}, M. 2012, \mnras, 426, 1808,
  \dodoi{10.1111/j.1365-2966.2012.21793.x}

\bibitem[{{Tinker} {et~al.}(2011){Tinker}, {Wetzel}, \& {Conroy}}]{Tinker2011}
{Tinker}, J., {Wetzel}, A., \& {Conroy}, C. 2011, arXiv e-prints,
  arXiv:1107.5046, \dodoi{10.48550/arXiv.1107.5046}

\bibitem[{{Wetzel} {et~al.}(2012){Wetzel}, {Tinker}, \& {Conroy}}]{Wetzel2012}
{Wetzel}, A.~R., {Tinker}, J.~L., \& {Conroy}, C. 2012, \mnras, 424, 232,
  \dodoi{10.1111/j.1365-2966.2012.21188.x}

\bibitem[{{Yang} {et~al.}(2022){Yang}, {Ianjamasimanana}, {Hammer}, {Higgs},
  {Namumba}, {Carignan}, {J{\'o}zsa}, \& {McConnachie}}]{Yang2022}
{Yang}, Y., {Ianjamasimanana}, R., {Hammer}, F., {et~al.} 2022, \aap, 660, L11,
  \dodoi{10.1051/0004-6361/202243307}

\end{thebibliography}





\end{document}